\documentclass[prc,amsmath,onecolumn,floatfix,superscriptaddress,showpacs,showkeys,nofootinbib,preprint]{revtex4}
\usepackage{epsfig}
\newcommand{\eq}[1]{\begin{align} #1 \end{align}}

\begin{document}

\title{Fluctuations of the $K/\pi$ Ratio in Nucleus-Nucleus Collisions:\\
Statistical and Transport Models}

\author{M.I.~Gorenstein}
\affiliation{Frankfurt Institute for Advanced Studies, Frankfurt, Germany}
\affiliation{Bogolyubov Institute for Theoretical Physics, Kiev, Ukraine}
\author{M.~Hauer}
\affiliation{Helmholtz Research School, University of Frankfurt, Frankfurt, Germany}
\author{V.P.~Konchakovski}
\affiliation{Bogolyubov Institute for Theoretical Physics, Kiev, Ukraine}
\affiliation{Helmholtz Research School, University of Frankfurt, Frankfurt, Germany}
\author{E.L.~Bratkovskaya}
\affiliation{Frankfurt Institute for Advanced Studies, Frankfurt, Germany}

\begin{abstract}

Event-by-event fluctuations of the kaon to pion number ratio in
nucleus-nucleus collisions are studied within the statistical
hadron-resonance gas model (SM) for different statistical ensembles and
in the Hadron-String-Dynamics (HSD) transport approach. We find that
the HSD model can qualitatively reproduce the measured excitation
function for the $K/\pi$ ratio fluctuations in central Au+Au (or Pb+Pb)
collisions from low SPS up to top RHIC energies. Substantial
differences in the HSD and SM results are found for the fluctuations
and correlations of the kaon and pion numbers. These predictions impose
a challenge for future experiments.
\end{abstract}


\maketitle


\section{Introduction}

The study of event-by-event fluctuations in high energy
nucleus-nucleus (A+A) collisions opens new possibilities to
investigate the phase transition between hadronic and partonic matter
as well as the QCD critical point (cf. the reviews \cite{fluc1}).
By measuring the fluctuations one might observe anomalies from the
onset of deconfinement~\cite{ood} and dynamical instabilities when the expanding
system goes through the 1-st order transition line between the
quark-gluon plasma and the hadron gas \cite{fluc2}. Furthermore, the
QCD critical point may be signaled by a characteristic pattern in
the fluctuations as pointed out in Ref. \cite{fluc3}.
However only recently, due to a rapid development of experimental
techniques, first measurements of the event-by-event fluctuations of
particle multiplicities \cite{NA49-1,fluc-mult,NA49-2,STAR} and
transverse momenta \cite{fluc-pT} in nucleus-nucleus collisions have
been performed.

From the theoretical side such event-by-event fluctuations for charged
hadron multiplicities (in nucleus-nucleus collisions) have been
studied in statistical models
\cite{CE,CE1,BF,res,MCE,CLT,acc,alpha,volume,power} and in dynamical
transport approaches \cite{HSD-1,HSDpp,urqmdpp,vokaSHINE}, which have
been used as important tools to investigate high-energy nuclear
collisions. We recall that the statistical models reproduce the mean
multiplicities of the produced hadrons (see e.g. Refs.
\cite{stat-model,FOC,FOP}), whereas the transport models (see, e.g.,
Refs.~\cite{HSD,UrQMD,transport}) provide, in addition, a dynamical
description of the various bulk properties of the system. By studying
the various fluctuations within statistical and transport models we
have found out that fluctuations provide an extremely sensitive
observable - depending on the details of the models - which are partly
washed out by looking at general quantities such as ensemble averages.

In particular, there is a qualitative difference in the properties of
the mean multiplicity and the scaled variance of the multiplicity
distribution in statistical models. In the case of mean multiplicities
the results obtained within the grand canonical ensemble (GCE), canonical
ensemble (CE), and micro-canonical ensemble (MCE) approach each other
in the large volume limit. One refers here to the thermodynamical
equivalence of the statistical ensembles. However, it was recently found
\cite{CE,MCE} that corresponding results for the scaled variances are
different in the GCE, CE and MCE ensembles, and thus the scaled variance is
sensitive to global conservation laws obeyed by a statistical system.
These differences are preserved in the thermodynamic limit.

Also there is a qualitative difference in the behavior of the
scaled variances of multiplicity distributions in statistical and
transport models. The transport models predict
\cite{HSDpp,urqmdpp} that the scaled variances in central
nucleus-nucleus collisions remain close to the corresponding
values in proton-proton collisions and increase with collision
energy in the same way as the corresponding multiplicities,
whereas in the statistical models the scaled variances approach
finite values at high collision energy, i.e. become independent of
energy. Accordingly, the differences in the scaled variance of
charged hadrons can be about factor of 10 at the top RHIC energy
\cite{HSDpp}. Only upcoming experimental data can clarify the
situation.

The QGP stage may form a specific set of primordial
fluctuation signals. A well known example is the equilibrium electric
charge fluctuation in QGP which is about a factor 2-3 smaller than
in an equilibrium hadron gas \cite{Heinz,Koch1}. To observe
primordial QGP fluctuations they should be frozen out during
expansion, hadronization, and further hadron-hadron
re-scatterings. Evolution and survival of the conserved charge
fluctuations in systems formed in nucleus-nucleus collisions at
the SPS and RHIC energies were discussed in
Refs.~\cite{Shuryak,Nayak}. Note that both the statistical models
and the HSD approach used in our study do not include the quark-gluon
degrees of freedom. Thus, the fluctuations in the QGP are outside
of the scope of the present paper.

The measurement of the fluctuations in the kaon to pion ratio by
the NA49 Collaboration \cite{NA49-1} was the first event-by-event
measurement in nucleus-nucleus collisions. It was suggested that
this ratio might allow to distinguish the enhanced strangeness
production attributed to the QGP phase.  Nowadays, the excitation
function for this observable is available in a wide range of
energies: from the NA49 collaboration  in Pb+Pb collisions at the
CERN SPS \cite{NA49-2}  and from the STAR collaboration in Au+Au
collisions at RHIC \cite{STAR}.  First statistical model estimates of
the $K/\pi$ fluctuations have been reported in Refs.~\cite{BH,Koch},
and results from the transport model UrQMD in Ref.~\cite{Ble}.

In this paper we present a systematic study of statistical model
results (in different ensembles) in comparison to HSD transport model
results for the fluctuations in the kaon to pion number ratio. The
paper is organized as follows: In Section II the characteristic
definitions for fluctuations in particle number ratios are introduced.
In Section III the relevant formulas of the statistical models (in
different ensembles) are presented. Statistical and HSD model results
for the fluctuations in the kaon to pion ratio for central
nucleus-nucleus collisions are compared in Section IV. In Section V the
HSD transport model results are additionally confronted with the
available data on $K/\pi$ fluctuations. A summary closes the paper in
Section VI.


\section{Measures of Particle Ratio Fluctuations}

\subsection{Notations and Approximations}

Let us introduce some notations.  We define the deviation $\Delta N_A$
from the average number $\langle N_A\rangle$ of the particle species
$A$ by $N_A=\langle N_A\rangle +\Delta N_A$.  Then we define covariance
for species $A$ and $B$
\eq{ \label{cov}
\Delta \left(N_A,N_B\right)~\equiv~\langle \Delta N_A
\Delta N_B\rangle~=~\langle N_A N_B\rangle
~-~\langle N_A\rangle \langle N_B\rangle~,
}
scaled variance
\eq{\label{omega}
\omega_A ~\equiv~\frac{\Delta\left(N_A,N_A\right)}{\langle
N_A\rangle}~=~\frac{\langle \left(\Delta N_A\right)^2\rangle}{\langle
N_A\rangle}~=~ \frac{\langle N_A^2\rangle -\langle
N_A\rangle^2}{\langle N_A\rangle}~,
}
and correlation coefficient
\eq{\label{corcoef}
\rho_{AB}~\equiv~\frac{\langle\Delta N_A~\Delta N_B\rangle}{\left[\langle\left(\Delta
N_A\right)^2\rangle~\langle\left(\Delta N_B\right)^2\rangle\right]^{1/2}}~.
}
The fluctuations of the ratio $R_{AB}\equiv N_A/N_B$
will be characterised by \cite{BH,Koch}
\eq{\label{sigma-def}
\sigma^2~\equiv~\frac{\langle \left(\Delta R_{AB}\right)^2\rangle}
{\langle R_{AB}\rangle^2 } ~.
}
Using the expansion,
\eq{\label{expand}
\frac{N_A}{N_B}~=~\frac{\langle N_A\rangle +\Delta N_A} {\langle
N_B \rangle +\Delta N_B}~ = ~ \frac{\langle N_A\rangle +\Delta
N_A} {\langle N_B \rangle }~ \times ~\left[1~-~\frac{\Delta N_B}{
\langle N_B \rangle } ~+~ \left(\frac{\Delta N_B}{ \langle N_B
\rangle}\right)^2~ -~\cdots~\right]~,
}
one finds to second order in $\Delta N_A/\langle N_A \rangle$ and
$\Delta N_B/\langle N_B \rangle$
the average value and the fluctuations of the $A$ to $B$ ratio:
\eq{\label{NANBav}
\langle R_{AB}\rangle~& \cong ~ \frac{\langle N_A\rangle}{\langle
N_B \rangle}~\left[1~+~ \frac{\omega_{B}}{\langle N_B
\rangle}~-~\frac{
\Delta\left(N_A,N_B\right)}{\langle N_A\rangle \langle
N_B\rangle}\right]~,
\\
\label{sigma}
 \sigma^2~&
\cong~\frac{\Delta \left(N_A,N_A\right)}{\langle
N_A\rangle^2}~+~\frac{\Delta \left(N_B,N_B\right)}{\langle
N_B\rangle^2}~-~2~\frac{\Delta \left(N_A,N_B\right)}{\langle
N_A\rangle\langle N_B\rangle}~
\nonumber \\
 &=~\frac{\omega_A}{\langle N_A\rangle}~+~\frac{\omega_B }{\langle
N_B\rangle}~-~2\rho_{AB}~\left[\frac{\omega_A \omega_B}{\langle
N_A\rangle\langle N_B\rangle}\right]^{1/2}~.
}
If species $A$ and $B$ fluctuate independently according to Poisson
distributions (this takes place, for example, in the GCE for an ideal
Boltzmann gas) one finds $\omega_A=\omega_B=1$ and $\rho_{AB}=0$.
Equation (\ref{sigma}) then reads
\eq{\label{Boltz}
\sigma^2~=~\frac{1}{\langle N_A\rangle}~+~\frac{1}{\langle
N_B\rangle}~.
}
In a thermal gas, the average multiplicities are proportional to the
system volume $V$.  Equation (\ref{Boltz}) demonstrates then a
simple dependence of $\sigma^2\propto 1/V$ on the system volume.

A few examples concerning to Eq.~(\ref{sigma}) are appropriate
here. When $\langle N_B\rangle \gg \langle N_A\rangle$, e.g.,
$A=K^++K^-$ and $B=\pi^++\pi^-$, the quantity $\sigma^2$
(\ref{sigma}) is dominated by the fluctuations of less
abundant particles. When $\langle N_A\rangle \cong \langle
N_B\rangle$, e.g., $A=\pi^+$ and $B=\pi^-$, the correlation term
in Eq.~(\ref{sigma}) may become especially important. A resonance
decaying always into a $\pi^+\pi^-$-pair does not contribute to
$\sigma^2$ (\ref{sigma}), but contributes to the $\pi^+$ and
$\pi^-$ average multiplicities. This leads \cite{Koch} to a
suppression of $\sigma^2$ (\ref{sigma}) in comparison to its value
given by Eq.~(\ref{Boltz}). For example, if all $\pi^+$ and
$\pi^-$ particles come in pairs from the decay of resonances, one
finds the correlation coefficient $\rho_{\pi^+\pi^-}=1$ in
Eq.~(\ref{sigma}), and thus $\sigma^2=0$. In this case, the
numbers of $\pi^+$ and $\pi^-$ fluctuate as the number of
resonances, but the ratio $\pi^+/\pi^-$ does not fluctuate.

\subsection{Mixed Events Procedure}

The experimental data for $N_A/N_B$ fluctuations are usually presented
in terms of the so called dynamical fluctuations
\cite{VKR}\footnote{Other dynamical measures, such as $\Phi$ \cite{GM,M} and
$F$ \cite{Koch}, can be also used.}
\eq{\label{sigmadyn}
\sigma_{dyn}~\equiv~\texttt{sign}\left(\sigma^2~-
~\sigma^2_{mix}\right)\left|\sigma^2~-~\sigma^2_{mix}\right|^{1/2}~,
}
where $\sigma^2$ is defined by Eq.~(\ref{sigma}) and $\sigma^2_{mix}$
corresponds to the following {\it mixed events} procedure\footnote{We
describe the idealized mixed events procedure appropriate for the model
analysis. The real experimental mixed events procedure is more
complicated and includes experimental uncertainties, such as particle
identification etc.}. One takes a large number of nucleus-nucleus
collision events and measures the numbers of $N_A$ and $N_B$ in each
event. Then all $A$ and $B$ particles from all events are combined into
one {\it set}. The construction of {\it mixed events} is done as
follows: One fixes a random number $N=N_A+N_B$ according to the
experimental probability distribution $P(N)$, takes randomly $N$
particles ($A$ and/or $B$) from the {\it whole set}, fixes the values
of $N_A$ and $N_B$, and returns these $N$ particles into the {\it set}.
This is the mixed event number one. Then one constructs event number 2,
number 3, etc.

Note that the number of events is much larger than the number of
hadrons, $N$, in any single event. Therefore, the probabilities
$p_{A}$ and $p_B = 1-p_{A}$, to take the $A$ and $B$ species from
the whole {\it set}, can be considered as constant values during
the event construction.  Another consequence of a large number of
events is the fact that $A$ and $B$ particles in any
constructed {\it mixed event} belong to different {\it physical
events} of nucleus-nucleus collisions. Therefore, the
correlations between the $N_B$ and $N_A$ numbers in a physical
event are expected to be destroyed in a mixed event. This is the
main purpose of the mixed events construction. For any function
$f(N_A,N_B)$ the mixed events averaging is then defined as
\begin{equation}\label{mix1}
\langle
f(N_A,N_B) \rangle_{mix} = \sum_N P(N) \sum_{N_A,N_B}
f(N_A,N_B) ~\delta(N-N_A-N_B) \frac{(N_A+N_B)!}{N_A! N_B!}
p_{A}^{N_A} p_B^{N_B}~.
\end{equation}
The straightforward calculations of mixed averages (\ref{mix1})
can be simplified by introducing the generating function $Z(x,y)$,
\eq{\label{mix2}
 & Z(x,y) ~\equiv~ \sum_N P(N) \sum_{N_A,N_B} \delta(N-N_A-N_B)
 \frac{(N_A +N_B)!}{N_A!~ N_B!}~ (xp_{A})^{N_A}~
 (yp_B)^{N_B}~
 \nonumber \\
& =~ \sum_N P(N) \left(xp_{A}+yp_{B}\right)^N~, }
which depends on auxiliary variables $x$ and $y$.
 The averages (\ref{mix1}) are then expressed as $x$- and
$y$-derivatives of $Z(x,y)$ at $x=y=1$.
One finds:
 \eq{ &\langle N_A \rangle_{mix} ~ = ~
 \left(\frac{\partial Z}{\partial x}\right)_{x=y=1}~ =~
 p_{A} ~\langle N \rangle ~,~~~~
 \langle N_B \rangle_{mix} ~ = ~
 \left(\frac{\partial Z}{\partial y}\right)_{x=y=1}~ = ~
 p_{B} ~ \langle N \rangle ~, \label{mix-av}\\
& \langle N_A(N_A-1) \rangle_{mix} ~ =~
 \left(\frac{\partial^2 Z}{\partial^2 x}\right)_{x=y=1} ~ =~
 p_{A}^2 ~\langle N (N-1)\rangle~,\label{mix-flA}\\~~~
& \langle N_B(N_B-1) \rangle_{mix} ~ =~
 \left(\frac{\partial^2 Z}{\partial^2 y}\right)_{x=y=1} ~ =~
 p_{B}^2 ~\langle N (N-1)\rangle~, \label{mix-flB}\\
&\langle N_AN_B \rangle_{mix} ~ - ~\langle N_A\rangle_{mix}
\langle N_B\rangle_{mix}~ = ~ \left(\frac{\partial^2 Z} {\partial
x\partial y}\right)_{x=y=1}
~=~ p_A p_B~\omega_N~
\langle N \rangle~~,\label{mix-cor}
 }
where
\eq{ \langle N \rangle ~\equiv ~\sum_N N~P(N)~,~~~~
\langle N^2 \rangle~\equiv ~\sum_N N^2~P(N)~,~~~~~
\omega_N~\equiv~
\frac{\langle N^2 \rangle~-~\langle N \rangle^2}{\langle N\rangle}~. \label{N-av-fluc}
}
Calculating the $N_A/N_B$ fluctuations for mixed events according
to Eq.~(\ref{sigma}) one
gets:
\eq{ & \sigma^2_{mix}~ \equiv~\frac{\Delta_{mix}
\left(N_A,N_A\right)}{\langle N_A\rangle^2}~+~\frac{\Delta_{mix}
\left(N_B,N_B\right)}{\langle N_B\rangle^2}~-~2~\frac{\Delta_{mix}
\left(N_A,N_B\right)}{\langle N_A\rangle\langle
N_B\rangle}~\nonumber \\
&=~\left[\frac{1}{\langle N_A \rangle }~+~ \frac{\omega_N-
1}{\langle N \rangle}\right] ~+~\left[\frac{1}{\langle N_B \rangle}~+~
\frac{\omega_N-1}{\langle N \rangle}\right]~-~
2~\frac{\omega_N-1}{\langle N \rangle}~\nonumber \\
&=~ \frac{1}{\langle N_A \rangle }~+~\frac{1}{\langle N_B \rangle
} ~. \label{Dmix}
}
A comparison of the final result in Eq.~(\ref{Dmix}) with
Eq.~(\ref{Boltz}) shows that the mixed events procedure gives the
same $\sigma^2$ for $N_A/N_B$ fluctuations as in the GCE
formulation for an ideal Boltzmann gas, i.e. for
$\omega_A=\omega_B=1$ and $\rho_{AB}=0$. If $\omega_N=1$ (e.g. for
the Poisson distribution $P(N)$), one indeed finds
$\omega_A^{mix}=\omega_B^{mix}=1$ and $\rho_{AB}^{mix}=0$.
Otherwise, if $\omega_N\neq 1$, the mixed events procedure leads
to $\omega_A^{mix}\neq 1$, $\omega_B^{mix}\neq 1$, and to non-zero
$N_AN_B$ correlations, as seen from the second line of
Eq.~(\ref{Dmix}).
Thus, if, e.g., event-by-event fluctuations in the total number of
pions and kaons are stronger than Poissonian ones, i.e. $\omega_N>1$,
positive pion-kaon correlations appear in the mixed events. They lead
to larger (smaller) $N_K$ in the sample of mixed events with larger
(smaller) $N_{\pi}$. However, the final result for $\sigma^2_{mix}$
(\ref{Dmix}) is still the same as for $\omega_N=1$, it does not depend
on the specific form of $P(N)$.
Non-trivial ($\omega_{A,B}^{mix}\neq 1$) fluctuations
of $N_A$ and $N_B$ as well as non-zero $\rho_{AB}^{mix}$
correlations may exist in the mixed events procedure, but they are
cancelled out in $\sigma_{mix}^2$.

\section{Fluctuations of Ratios in Statistical Models}

\subsection{Quantum Statistics and Resonance Decays}

The occupation numbers, $n_{{\bf p},j}$, of single quantum states
(with fixed projection of particle spin) labelled by the momentum
vector ${\bf p}$ are equal to $n_{{\bf p},j}=0,1,\ldots,\infty$
for bosons and $n_{{\bf p},j}=0,1$ for fermions. Their average
values are
 \eq{
 \langle n_{{\bf p},j} \rangle
 ~& = ~\frac {1} {\exp \left[\left( \epsilon_{{\bf p}j} - \mu_j \right)/ T\right]
 ~-~ \alpha_j}~, \label{np-aver}
 }
and their fluctuations read
\eq{
 \langle~\left(\Delta n_{{\bf p},j}\right)^2~\rangle_{gce}
~ \equiv ~ \langle \left( n_{{\bf p},j}~-~\langle n_{{\bf
p},j}\rangle\right)^2\rangle_{gce} ~=~ \langle n_{{\bf p},j}\rangle
\left(1~ + ~\alpha_j ~\langle n_{{\bf p},j}
\rangle\right)~\equiv~v^{ 2}_{{\bf p},j}~,
\label{np-fluc}
}
where $T$ is the system temperature, $m_j$ is the mass of a
particle $j$, $\epsilon_{{\bf p}j}=\sqrt{{\bf p}^{2}+m_j^{2}}$ is
the single particle energy. The value of $\alpha_j$ depends on
quantum statistics, i.e. $+1$ for bosons and $-1$ for fermions,
while $\alpha_j=0$ gives the Boltzmann approximation. The chemical
potential $\mu_j$ of a species $j$ equals to:
$\mu_j~=~q_j~\mu_Q~+~b_j~\mu_B~+~s_j~\mu_S $,
where $q_j,~b_j,~s_j$ are the particle electric charge, baryon
number, and strangeness, respectively, while $\mu_Q,~\mu_B,~\mu_S$
are the corresponding chemical potentials which regulate the
average values of these global conserved charges in the GCE.

In the equilibrium hadron-resonance gas model the mean
number of primary particles (or resonances) is
calculated as:
 \eq{\label{Ni-gce}
 \langle N_j^*\rangle \;\equiv\; \sum_{\bf p} \langle n_{{\bf p},j}\rangle
 \;=\; \frac{g_j V}{2\pi^{2}}\int_{0}^{\infty}p^{2}dp\; \langle
 n_{{\bf p},j}\rangle\;,
}
where $V$ is the system volume and $g_j$ is the degeneracy factor
of a particle of species $j$ (the number of spin states). In the
thermodynamic limit, $V\rightarrow \infty$, the sum over the
momentum states can be substituted by a momentum integral.

It is convenient to introduce a microscopic correlator, $\langle
\Delta n_{{\bf p},j} \Delta n_{{\bf k},i} \rangle$, which in the
GCE has the simple form:
\eq{ \label{mcc-gce}
\langle \Delta n_{{\bf p},j}~ \Delta n_{{\bf k},i}
\rangle_{gce}~=~ \upsilon_{{\bf
p},j}^2\,\delta_{ij}\,\delta_{{\bf p}{\bf k}}~.
}
Hence there are no correlations between different particle
species, $i\neq j$, and/or between different momentum states,
${\bf p} \neq {\bf k}$. Only the Bose enhancement, $v_{{\bf
p},j}^2>\langle n_{{\bf p},j}\rangle$ for $\alpha_j=1$, and the
Fermi suppression, $v_{{\bf p},j}^2<\langle n_{{\bf p},j}\rangle$
for $\alpha_j=-1$, exist for fluctuations of primary particles in
the GCE. The correlator (\ref{cov}) can be presented
in terms of microscopic correlators (\ref{mcc-gce}):
\eq{ \label{dNidNj}
\langle \Delta N_j^* ~\Delta N_i^*~\rangle_{gce} ~=~ \sum_{{\bf
p},{\bf k}}~\langle \Delta n_{{\bf p},j}~\Delta n_{{\bf
k},i}\rangle_{gce}~=~\delta_{ij}~\sum_{\bf p}~ v_{{\bf
p},j}^2~.
 }
In the case $i=j$  equation (\ref{dNidNj}) gives the variance
of primordial particles (before resonance decays) in the GCE.

For the hadron resonance gas formed in relativistic A+A collisions the
corrections due to quantum statistics (Bose enhancement and Fermi
suppression) are small\footnote{Possible strong Bose effects are
discussed in Ref.~\cite{BF}}. For the pion gas at $T=160$~MeV, one
finds $\omega_{\pi}\cong 1.1$, instead of $\omega=1$ for Boltzmann
particles. The quantum statistics effects are even smaller for heavier
particles like kaons and almost negligible for resonances.

The average final (after resonance decays) multiplicities
$\langle N_i\rangle$ are equal to:
 \eq{\label{<N>}
 \langle N_i\rangle
 \;=\;
 \langle N_i^*\rangle + \sum_R \langle N_R\rangle \langle
 n_{i}\rangle_R\;.
 }
In Eq.~(\ref{<N>}), $N_i^*$ denotes the number of stable primary
hadrons of species $i$, the summation $\sum_R$ runs over all types
of resonances $R$, and $\langle n_i\rangle_R \equiv \sum_r b_r^R
n_{i,r}^R$~ is the average over resonance decay channels. The
parameters $b^R_r$ are the branching ratios of the $r$-th
branches, $n_{i,r}^R$ is the number of particles of species $i$
produced in resonance $R$ decays via a decay mode $r$. The index
$r$ runs over all decay channels of a resonance $R$ with the
requirement $\sum_{r} b_r^R=1$.
In the GCE the correlator (\ref{cov}) after resonance decays
can be calculated as
\cite{Koch}:
 \eq{\label{corr-GCE}
 \langle \Delta N_A \Delta N_B\rangle_{gce}
 =
 \langle\Delta N_A^* \Delta N_B^*\rangle_{gce}
 + \sum_R \left[ \langle\Delta N_R^2\rangle
 \langle n_{A}\rangle_R \langle n_{B}\rangle_R
 + \langle N_R\rangle \langle \Delta n_{A}\Delta n_{B}\rangle_R
 \right]~,
 }
where $\langle \Delta n_A~\Delta n_B\rangle_R\equiv \sum_r b_r^R
n_{A,r}^R n_{B,r}^R~-~\langle n_A\rangle_R\langle n_B\rangle_R$~.

\subsection{Global Conservation Laws}

In the MCE, the energy and conserved charges are fixed exactly for
each microscopic state of the system. This leads to two
modifications in comparison with the GCE. First, additional
terms appear for the primordial microscopic correlators in the
MCE. They reflect the (anti)correlations between different
particles, $i\neq j$, and different momentum levels, ${\bf p}\neq
{\bf k}$, due to charge and energy conservation in the MCE \cite{MCE},
 \eq{\label{corr}
 &
 \langle \Delta n_{{\bf p},j} \Delta n_{{\bf k},i} \rangle_{mce}
 ~=\; \upsilon_{{\bf p},j}^2\,\delta_{ij}\,\delta_{{\bf p}{\bf k}}
 \;-\; \frac{\upsilon_{{\bf p},j}^2v_{{\bf k},i}^2}{|A|}\;
 [\;q_iq_j M_{qq} + b_ib_j M_{bb} + s_is_j M_{ss} \nonumber
 \\
 &+ ~\left(q_is_j + q_js_i\right) M_{qs}~
 - ~\left(q_ib_j + q_jb_i\right) M_{qb}~
 - ~\left(b_is_j + b_js_i\right) M_{bs}\nonumber
 \\
 &+~ \epsilon_{{\bf p}j}\epsilon_{{\bf k}i} M_{\epsilon\epsilon}~-~
 \left(q_i \epsilon_{{\bf p}j} + q_j\epsilon_{{\bf k}i} \right)
 M_{q\epsilon}~
 +~ \left(b_i \epsilon_{{\bf p}j} + b_j\epsilon_{{\bf k}i} \right)
 M_{b\epsilon}~
 - ~\left(s_i \epsilon_{{\bf p}j} + s_j\epsilon_{{\bf k}i} \right) M_{s\epsilon}
 \;]\;,
 }
where $|A|$ is the determinant and $M_{ij}$ are the minors of the
following matrix,
 \eq{\label{matrix}
 A =
 \begin{pmatrix}
 \Delta (q^2) & \Delta (bq) & \Delta (sq) & \Delta (\epsilon q)\\
 \Delta (q b) & \Delta (b^2) & \Delta (sb) & \Delta (\epsilon b)\\
 \Delta (q s) & \Delta (b s) & \Delta (s^2) & \Delta (\epsilon s)\\
 \Delta (q \epsilon) & \Delta (b \epsilon) & \Delta (s \epsilon) & \Delta (\epsilon^2)
 \end{pmatrix}\;,
 }
with the elements, $\;\Delta (q^2)\equiv\sum_{{\bf p},j}
q_{j}^2\upsilon_{{\bf p},j}^2\;$, $\;\Delta (qb)\equiv \sum_{{\bf
p},j} q_{j}b_{j}\upsilon_{{\bf p},j}^2\;$, $\;\Delta
(q\epsilon)\equiv \sum_{{\bf p},j} q_{j}\epsilon_{{\bf
p}j}\upsilon_{{\bf p},j}^2\;$, etc. The sum, $\sum_{{\bf p},j}$~,
means integration over momentum ${\bf p}$, and the summation over
all hadron-resonance species~$j$ contained in the model. The first
term in the r.h.s. of Eq.~(\ref{corr}) corresponds to the
microscopic correlator (\ref{mcc-gce}) in the GCE. Note, that the presence
of the terms containing the single particle energy $\epsilon_{{\bf
p}j}=\sqrt{{\bf p}^{2}+m_j^{2}}$ in Eq.~(\ref{corr}) is a consequence
of energy conservation. In the CE, only charges are conserved, thus
the terms containing $\epsilon_{{\bf p}j}$ in Eq.~(\ref{corr}) are
absent. The matrix $A$ in Eq.~(\ref{matrix}) then becomes a $3\times 3$
matrix (see Ref.~\cite{res}). An important property of the microscopic
correlator method is that the particle number fluctuations and the
correlations in the MCE or CE, although being different from those in
the GCE, are expressed by quantities calculated within the GCE. The
microscopic correlator (\ref{corr}) can be used to calculate the
primordial particle (or resonances) correlator in the MCE (or in the
CE):
\eq{
 \langle \Delta N_{i} ~\Delta N_{j}~\rangle_{mce}
 &~= \sum_{{\bf p},{\bf k}}~\langle \Delta n_{{\bf p},i}~\Delta
 n_{{\bf k},j}\rangle_{mce}\;. \label{mc-corr-mce}
}

A second feature of the MCE (or CE) is the modification of the
resonance decay contribution to the fluctuations in comparison to
the GCE (\ref{corr-GCE}). In the MCE (or CE) it reads\cite{res,MCE}:
\eq{
 \langle \Delta N_A\,\Delta N_B\rangle_{mce}
 ~&=\; \langle\Delta N_A^* \Delta N_B^*\rangle_{mce}
 \;+\; \sum_R \langle N_R\rangle\; \langle \Delta n_{A}\; \Delta n_{B}\rangle_R
 \;+\; \sum_R \langle\Delta N_A^*\; \Delta N_R\rangle_{mce}\; \langle n_{B}\rangle_R
 \; \nonumber
 \\
 &+\; \sum_R \langle\Delta N_B^*\;\Delta N_R\rangle_{mce}\; \langle n_{A}\rangle_R
 \;+\; \sum_{R, R'} \langle\Delta N_R\;\Delta N_{R'}\rangle_{mce}
 \; \langle n_{A}\rangle_R\;
 \langle n_{B}\rangle_{R^{'}}\;.\label{corr-MCE}
 }
Additional terms in Eq.~(\ref{corr-MCE}) compared to
Eq.~(\ref{corr-GCE}) are due to the correlations (for primordial
particles) induced by energy and charge conservations in the MCE.
Eq.~(\ref{corr-MCE}) has the same form in the CE \cite{res}
and MCE \cite{MCE}, the difference between these two ensembles appears
because of different microscopic correlators (\ref{corr}). The
microscopic correlators of the MCE together with
Eq.~(\ref{mc-corr-mce}) should be used to calculate
 the correlators $\langle\Delta N_A^*
\Delta N_B^*\rangle_{mce}$~,~$\langle\Delta N_A^*\; \Delta
N_R\rangle_{mce}~$, $~\langle\Delta N_A^*\;\Delta
N_R\rangle_{mce}~$, $~\langle\Delta N_B^*\;\Delta
N_R\rangle_{mce}~$, and $~\langle\Delta N_R\;\Delta
N_{R'}\rangle_{mce}$ entering in Eq.~(\ref{corr-MCE}).
The correlators (\ref{corr-MCE}) define finally
the scaled variances $\omega_A$
and $\omega_B$ (\ref{omega}) and
correlations $\rho_{AB}$ (\ref{corcoef}) between the $N_A$ and $N_B$
numbers. Together with the average multiplicities $\langle N_A\rangle$
and $\langle N_B\rangle$ they define completely the fluctuations
$\sigma^2$ (\ref{sigma}) of the $A$ to $B$ number ratio.

\section{Statistical and HSD model results for the $K/\pi$ ratio}

In this section we present the results of the hadron-resonance gas
statistical model (SM) and the HSD transport model for the
fluctuations of the $K/\pi$ ratio in central nucleus-nucleus
collisions. To carry out the SM calculations one has to fix the
chemical freeze-out parameters. The dependence of the chemical
potential $\mu_B$ on the collision energy is parameterized as
\cite{FOC}:
$\mu_B \left( \sqrt{s_{NN}} \right) =1.308~\mbox{GeV}\cdot(1+
0.273~ \sqrt{s_{NN}})^{-1}~,$
where the center of mass nucleon-nucleon collision energy,
$\sqrt{s_{NN}}$, is taken in units of GeV. The system is assumed
to be net strangeness free, i.e. $S=0$, and to have the charge to
baryon ratio of the initial colliding nuclei, i.e. $Q/B = 0.4$.
These two conditions define the system strange, $\mu_S$, and
electric, $\mu _Q$, chemical potentials. For the chemical
freeze-out condition we chose the average energy per particle,
$\langle E \rangle/\langle N \rangle = 1~$GeV \cite{Cl-Red}.
Finally, the strangeness saturation factor, $\gamma_S$, is
parameterized as in Ref. \cite{FOP}:
$ \gamma_S~ =~ 1 - 0.396~ \exp \left( - ~1.23~ T/\mu_B \right). $
This determines all parameters of the model. An extended version
of the THERMUS framework \cite{Thermus} is used for the SM
calculations (for more details see Ref.~\cite{MCE}).
Note that average multiplicities of pions and kaons measured
in central nucleus-nucleus collisions at  SPS and RHIC energies
\cite{Kpi-mult} are nicely described in the SM (see, e.g.,
Refs.~\cite{FOC,FOP}) as well as HSD (cf. Ref.~\cite{transport}).

\subsection{Results for $\omega_K$, $\omega_{\pi}$, and $\rho_{K\pi}$}

According to Eq.~(\ref{sigma}) the fluctuations of the $K=K^++K^-$ to $\pi=\pi^++\pi^-$
ratio is given by
\eq{
 \sigma^2~
 =~\frac{\omega_K}{\langle N_K\rangle}~+~\frac{\omega_{\pi} }{\langle
N_{\pi}\rangle}~-~2\rho_{K\pi}~\left[\frac{\omega_K ~ \omega_{\pi} } {\langle
N_K\rangle\langle N_{\pi}\rangle}\right]^{1/2}~.
\label{sigmaKpi}
}
\begin{table}[ht!]
 \begin{center}
 {\footnotesize
 \begin{tabular}{||c||c|c||c|c||c|c|c||c|c|c||c|c|c||c|c|c||}\hline
 \ $\sqrt{s_{NN}}$ &\ $T$ &\ $\mu_B$ & $n_\pi$ &\ $n_K$ & \multicolumn{3}{c||}{GCE} & \multicolumn{3}{c||}{CE} & \multicolumn{3}{c||} {MCE} \\\hline
 \ [ GeV ] &\ [ MeV ] &\ [ MeV ] &\ $[ fm^{-3} ]$ &\ $[ fm^{-3} ]$ &\ $\omega_\pi$ &\ $\omega_K$ &\ $\rho_{K\pi}$ &\ $\omega_\pi$ &\ $\omega_K$ &\ $\rho_{K\pi}$ &\ $\omega_\pi$ &\ $\omega_K$ &\ $\rho_{K\pi}$\\\hline\hline
     \ 6.27&\ 130.7&\ 482.4&\ 0.106&\ 0.011 &\ 1.247&\ 1.030&\ 0.055&\ 1.122&\ 0.930&\ 0.038&\ 0.641&\ 0.833&\ -0.243\\
     \ 7.62&\ 138.3&\ 424.6&\ 0.134&\ 0.016 &\ 1.301&\ 1.039&\ 0.066&\ 1.184&\ 0.961&\ 0.049&\ 0.656&\ 0.853&\ -0.249\\
     \ 8.77&\ 142.9&\ 385.4&\ 0.155&\ 0.020 &\ 1.337&\ 1.045&\ 0.073&\ 1.228&\ 0.980&\ 0.057&\ 0.669&\ 0.866&\ -0.251\\
     \ 12.3&\ 151.5&\ 300.1&\ 0.202&\ 0.029 &\ 1.408&\ 1.058&\ 0.086&\ 1.324&\ 1.018&\ 0.074&\ 0.705&\ 0.893&\ -0.242\\
     \ 17.3&\ 157  &\ 228.6&\ 0.239&\ 0.038 &\ 1.457&\ 1.068&\ 0.095&\ 1.397&\ 1.044&\ 0.087&\ 0.743&\ 0.915&\ -0.226\\
     \ 62.4&\ 163.1&\ 72.7 &\ 0.293&\ 0.055 &\ 1.514&\ 1.084&\ 0.110&\ 1.506&\ 1.081&\ 0.109&\ 0.824&\ 0.947&\ -0.186\\
     \ 130 &\ 163.6&\ 36.1 &\ 0.298&\ 0.058 &\ 1.519&\ 1.086&\ 0.112&\ 1.516&\ 1.085&\ 0.111&\ 0.833&\ 0.950&\ -0.181\\
     \ 200 &\ 163.7&\ 23.4 &\ 0.298&\ 0.058 &\ 1.519&\ 1.086&\ 0.112&\ 1.519&\ 1.085&\ 0.112&\ 0.835&\ 0.950&\ -0.180\\
     \hline
 \end{tabular}
 }
 \caption{The chemical freeze-out parameters $T$ and $\mu_B$ for central Pb+Pb (Au+Au) collisions
at different c.m. energies $\sqrt{s_{NN}}$. The hadron-resonance gas model results are presented for
final (after resonance decays) number densities of pions $n_{\pi}$ and kaons $n_K$
(they are the same in all statistical ensembles),
scaled variances $\omega_{\pi}$, $\omega_K$ and correlation parameter $\rho_{K\pi}$
in the GCE, CE, and MCE.}
 \end{center}
\end{table}
\begin{table}[ht!]
 \begin{center}
 {\footnotesize
 \begin{tabular}{||c||c|c|c|c|c||}\hline
 \ $\sqrt{s_{NN}}$ & \multicolumn{5}{c||}{HSD full acceptance} \\\hline
 \ [ GeV ] &\ $\langle N_\pi\rangle $ &\ $\langle N_K\rangle
 $ &\ $\omega_\pi$ &\ $\omega_K$ &\ $\rho_{K\pi}$\\\hline\hline
     \ 6.27&\ 612.03&\ 43.329 &\ 0.961&\ 1.107&\ -0.091\\
     \ 7.62&\ 732.11&\ 60.801 &\ 1.077&\ 1.141&\ -0.063\\
     \ 8.77&\ 823.71&\ 75.133 &\ 1.159&\ 1.168&\ -0.033\\
     \ 12.3&\ 1072.3&\ 116.44 &\ 1.378&\ 1.250&\  0.046\\
     \ 17.3&\ 1364.6&\ 165.52 &\ 1.619&\ 1.348&\  0.126\\
     \ 62.4&\ 2933.9&\ 449.29 &\ 3.006&\ 1.891&\  0.412\\
     \ 130 &\ 4304.2&\ 692.59 &\ 4.538&\ 2.378&\  0.557\\
     \ 200 &\ 5204.0&\ 861.77 &\ 5.838&\ 2.765&\  0.634\\
 \hline
 \end{tabular}
 }
 \caption{The HSD average multiplicities $\langle N_{\pi}\rangle$,
 $\langle N_K\rangle$ and values of $\omega_{\pi}$, $\omega_K$, and
$\rho_{K\pi}$ for central (impact parameter $b=0$) Pb+Pb (Au+Au)
collisions at different c.m. energies $\sqrt{s_{NN}}$.}
 \label{Tab1_N}
 \end{center}
\end{table}
\begin{figure}[ht!]
 \epsfig{file=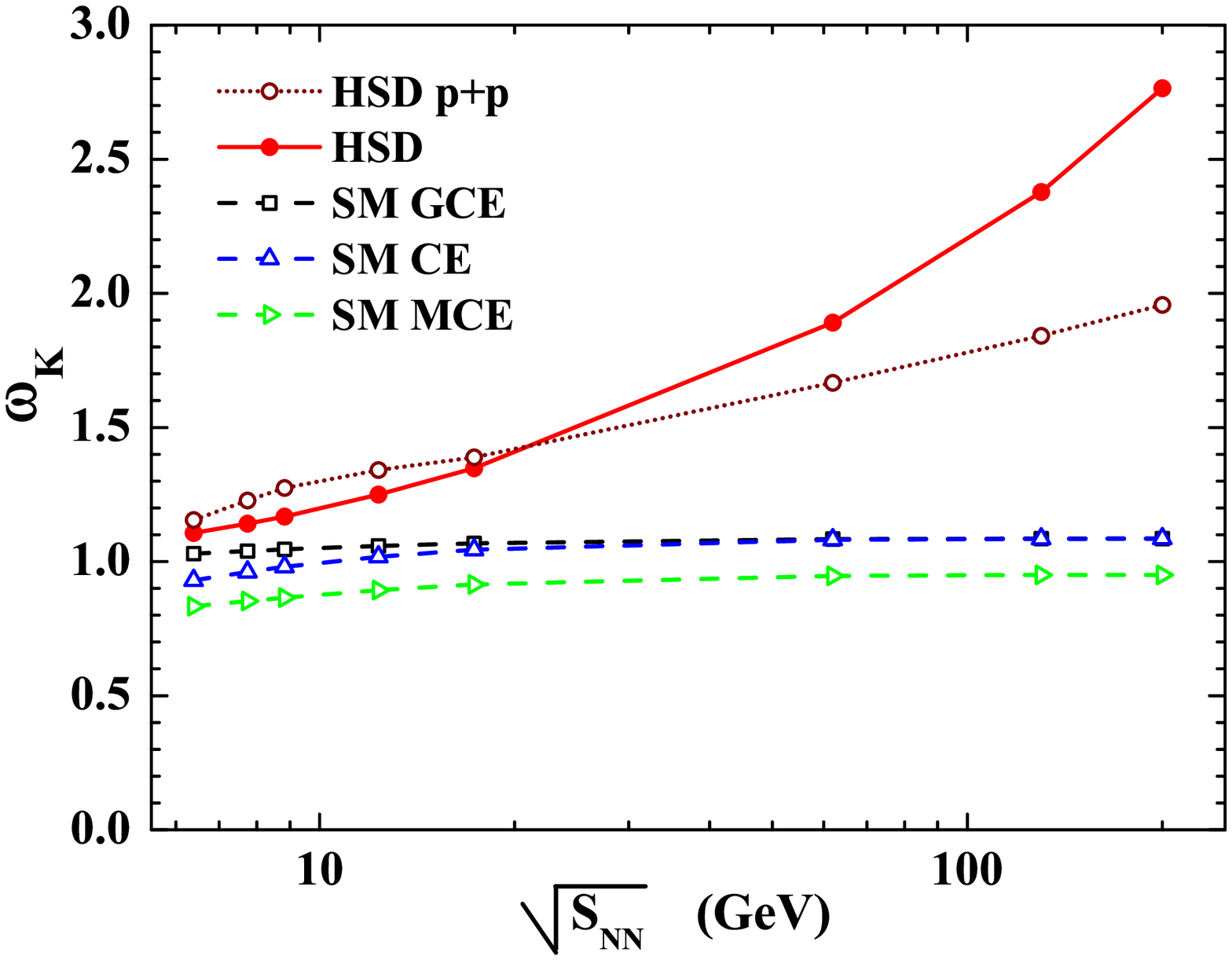,width=12cm}
 \epsfig{file=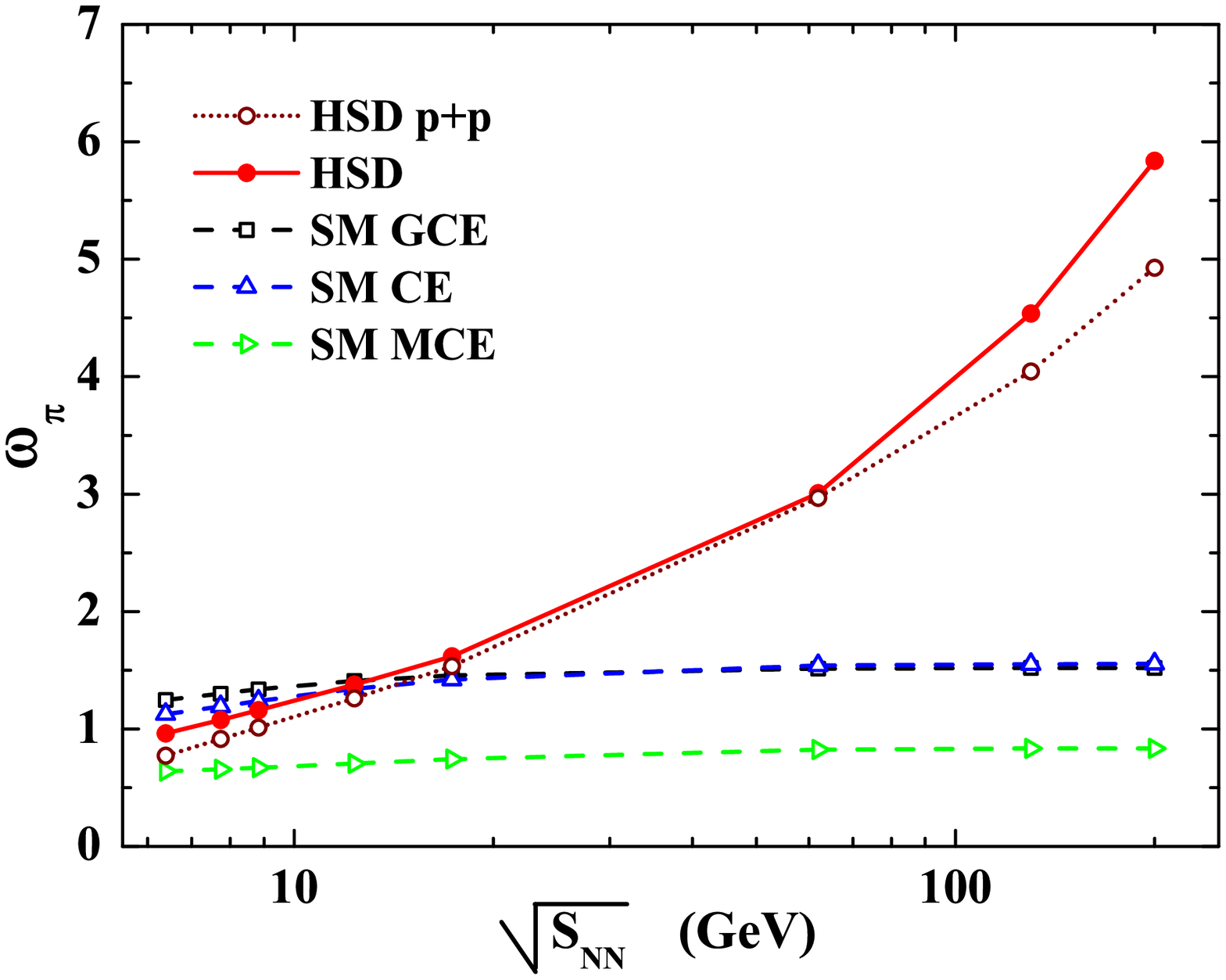,width=12cm}
 \caption{(Color online) The SM results in the GCE, CE, and MCE ensembles and the HSD
results (impact parameter $b=0$) are presented for the scaled variances
$\omega_{\pi}$, $\omega_K$ for Pb+Pb (Au+Au) collisions at different
c.m. energies $\sqrt{s_{NN}}$. For comparison the HSD results for
inelastic proton-proton collisions are also presented in terms of the
dotted  lines with open circles.}
\label{fig1}
\end{figure}
\begin{figure}[ht!]
 \epsfig{file=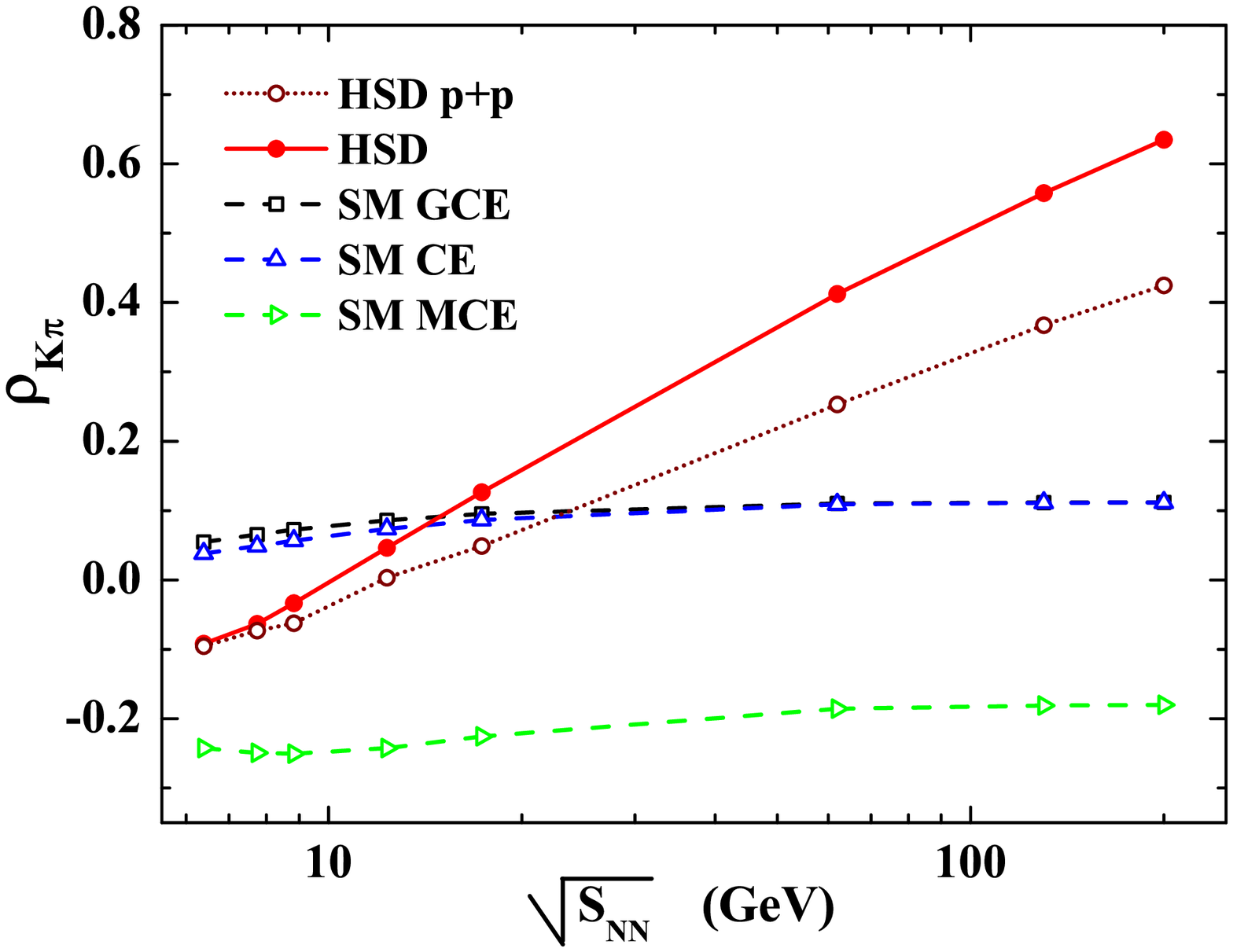,width=12cm}
\caption{(Color online) The SM results in the GCE, CE, and MCE ensembles and the HSD
results (impact parameter $b=0$) are presented for the correlation
parameter $\rho_{K\pi}$ for Pb+Pb (Au+Au) collisions at different c.m.
energies $\sqrt{s_{NN}}$. For comparison the HSD results for inelastic
proton-proton collisions are also presented
by the dotted line with open circles.}
\label{fig2}
\end{figure}
The values of $\omega_{\pi}$, $\omega_K$ and $\rho_{K\pi}$ in different
statistical ensembles are presented in Table I and for the HSD
simulations of Pb+Pb (Au+Au) central (with impact parameter $b=0$)
collisions in Table II. Both the SM and HSD results are shown in
Figs.~\ref{fig1} and \ref{fig2}. Let us first comment the SM results.
In the SM the scaled variances $\omega_{\pi}$ and $\omega_K$ and
correlation parameter $\rho_{K\pi}$ approach finite values in the
thermodynamic limit of large volumes. These limiting values are
presented in Table I and in Fig.~\ref{fig1} and \ref{fig2}. For
central Pb+Pb and Au+Au collisions the corresponding volumes in the SM are
large enough. Finite volume corrections are expected to be on the level of
a few percent. The finite volume effects for the scaled variances and
correlation parameters in the CE and MCE are, however, difficult to
calculate (see Ref.~\cite{CLT}) and they will not be considered in the
present paper. The GCE values of $\omega_{\pi}$ and $\omega_K$
reflect the Bose statistics of pions and kaons and the
contributions from resonance decays.

The $\pi$-$K$ correlations $\rho_{K\pi}$ are due to resonances having
simultaneously $K$ and $\pi$ mesons in their decay products. In the
hadron-resonance gas within the GCE ensemble, these quantum statistics
and resonance decay effects are responsible for deviations of
$\omega_K$ and $\omega_{\pi}$ from 1, and of $\rho_{K\pi}$ from 0. The
most important effect of an exact charge conservation in the CE
ensemble is a suppression of the kaon number fluctuation.  This happens
mainly due to exact strangeness conservation and is reflected in
smaller CE values of $\omega_K$ at low collision energies in comparison
to those from the GCE ensemble. The MCE values of $\omega_K$ and
$\omega_{\pi}$ are further suppressed in comparison those from the CE
ensemble, which is due to exact energy conservation. The effect is
stronger for pions than for kaons since pions carry a larger part of
the total energy.  An important feature of the MCE is the
anticorrelation between $N_{\pi}$ and $N_K$, i.e. negative values of
$\rho_{K\pi}$. This is also a consequence of energy conservation for
each microscopic state of the system in the MCE \cite{MCE}. The
presented results demonstrate that global conservation laws are rather
important for the values of $\omega_{\pi}$, $\omega_K$, and
 $\rho_{K\pi}$. In particular, the exact energy conservation strongly
suppresses the fluctuations in the pion and kaon numbers and leads to
$\omega_K<1$ and $\omega_{\pi}<1$ in the MCE ensemble instead of
$\omega_K>1$ and $\omega_{\pi}>1$ in the GCE and CE ensembles.  The
exact energy conservation changes also the $\pi$-$K$ correlation into
an anticorrelation: instead of $\rho_{K\pi}>0$ in the GCE and CE
ensembles one finds $\rho_{K\pi}<0$ in the MCE.

As seen from Figs.~\ref{fig1} and \ref{fig2} the HSD results for
$\omega_{\pi}$, $\omega_K$, and $\rho_{K\pi}$ (solid lines) are
rather different from those in the SM. For a comparison the HSD
results for inelastic proton-proton collisions are also presented
in Figs.~\ref{fig1} and \ref{fig2} (dotted lines). The HSD scaled
variances $\omega_{\pi}$ and $\omega_K$ increase at higher
energies. A similar behavior has been observed earlier in
Ref.~\cite{HSDpp} for the scaled variance of all charged hadrons.
The HSD calculations reveal the anticorrelation between $N_{\pi}$
and $N_K$, i.e. negative values of $\rho_{K\pi}$, for low SPS
energies, where the influence of conservation laws is more
stringent.

Comparing this result with the SM (in different ensembles) one may
conclude that negative values of $\rho_{K\pi}$ in  HSD appear
because of a dominant role of energy conservation in joint
$\pi$-$K$ production at small collision energies. The HSD values
of $\rho_{K\pi}$ become, however, positive and strongly increases
with increasing collision energy. This is due to the contribution
of heavy strings to joint $\pi$-$K$ (or $K^*$) production at high
energies in the HSD simulations. Note that the HSD results for
$\omega_{\pi}$, $\omega_K$, and $\rho_{K\pi}$ in nucleus-nucleus
collisions become larger than those in proton-proton inelastic
reactions at high collision energies. This is due to an increase
of secondary (i.e. meson-baryon and meson-meson) collisions at
higher bombarding energy. Thus, a strong deviation of HSD from the
SM with increasing energies is a consequence of non-equilibrium
dynamics in the hadron-string model which is driven by the
formation of heavy strings and their decay. Indeed, future
experimental data on the fluctuations of $K, \pi$ and
$K\pi$-correlations will allow to shed more light on the
equilibration pattern achieved in heavy-ion collisions at
RHIC energies.

Two comments are appropriate here. The first one concerns a
correspondence between the HSD and SM results.  In HSD three
charges -- net baryon number $B$ (equal to the number of
participating nucleons), net electric charge $Q$ (equal to the
number of participating protons), and net strangeness $S$ (equal to
zero) -- are conserved exactly during the system evolution.
However, $B$ and $Q$ can fluctuate from event to event because of
the fluctuations in the number of nucleon participants. They also
cause fluctuations of the energy of produced hadrons in the HSD
simulations. Besides, an essential part of the system energy is
transformed to collective motion. Thus, even in the sample of
the HSD events with fixed number of participants, the thermal
energy of the created particles can fluctuate from event to event.
Both the charge and energy fluctuations in HSD are not of
thermal origin. Therefore, an attempt to interpret the HSD
multiplicity fluctuations in statistical terms would
require to use a more general concept of statistical ensembles with
fluctuating extensive quantities \cite{alpha,volume}. In
particular, large values of the scaled variances, $\omega_i\sim
\langle N_i\rangle$, in high energy proton-proton collisions are
also present in the HSD simulations of nucleus-nucleus collisions.
In the SM model this would require a special form of scaling
volume fluctuations as  recently suggested in
Ref.~\cite{power}.

Our second comment concerns the physical origin of the correlation
parameter $\rho_{K\pi}$. Two sources of the $\pi$-$K$ correlations
are: resonance, string decays and electric charge conservation. To
estimate their relative weights, one can benefit from measuring
the correlations $\rho_{K\pi}$ in the separate charge channels:
$\pi^-K^-$ and $\pi^-K^+$ as  suggested in
Ref.~\cite{Mrow}. The resonances decaying into $\pi^-K^+$ produce
the corresponding correlation, while an analogous correlation in the
$\pi^-K^-$ system is absent. Note that electric charge conservation leads
also to qualitatively different correlation effects in $\pi^-K^-$
and $\pi^-K^+$ channels.

\subsection{Results for $\sigma$, $\sigma_{mix}$, and $\sigma_{dyn}$}

The fluctuation in the kaon to pion ratio is dominated by the
fluctuations of kaons alone since the average multiplicity of kaons is
about 10 times smaller than that of pions. Thus, the 1-st term in the
r.h.s. of Eq.~(\ref{sigmaKpi}) gives the dominant contribution, while the
2-nd and 3-rd terms in (\ref{sigmaKpi}) give only small corrections.
The model calculations of (\ref{sigmaKpi}) require, in addition to
$\omega_K$, $\omega_{\pi}$, and $\rho_{K\pi}$ values, the knowledge of
the average multiplicities $\langle N_K\rangle$ and $\langle
N_{\pi}\rangle$. For the HSD simulations (impact parameter $b=0$ in Pb+Pb
collisions at SPS energies and Au+Au collisions at RHIC) the
corresponding average multiplicities are presented in Table II. To fix
average multiplicities in the SM one needs to choose the system volume.
For each collision energy we fix the volume of the statistical system
to obtain the same kaon average multiplicity in the SM as in the
HSD calculations: $\langle N_K\rangle_{stat}= \langle
N_K\rangle_{HSD}$. We recall that average multiplicities of kaons and pions
are the same in all statistical ensembles. The SM volume in central
Pb+Pb (Au+Au) collisions is large enough and all statistical ensembles
are thermodynamically equivalent for the average pion and kaon
multiplicities since these multiplicities are much larger than 1.

\begin{figure}[h!]
  \epsfig{file=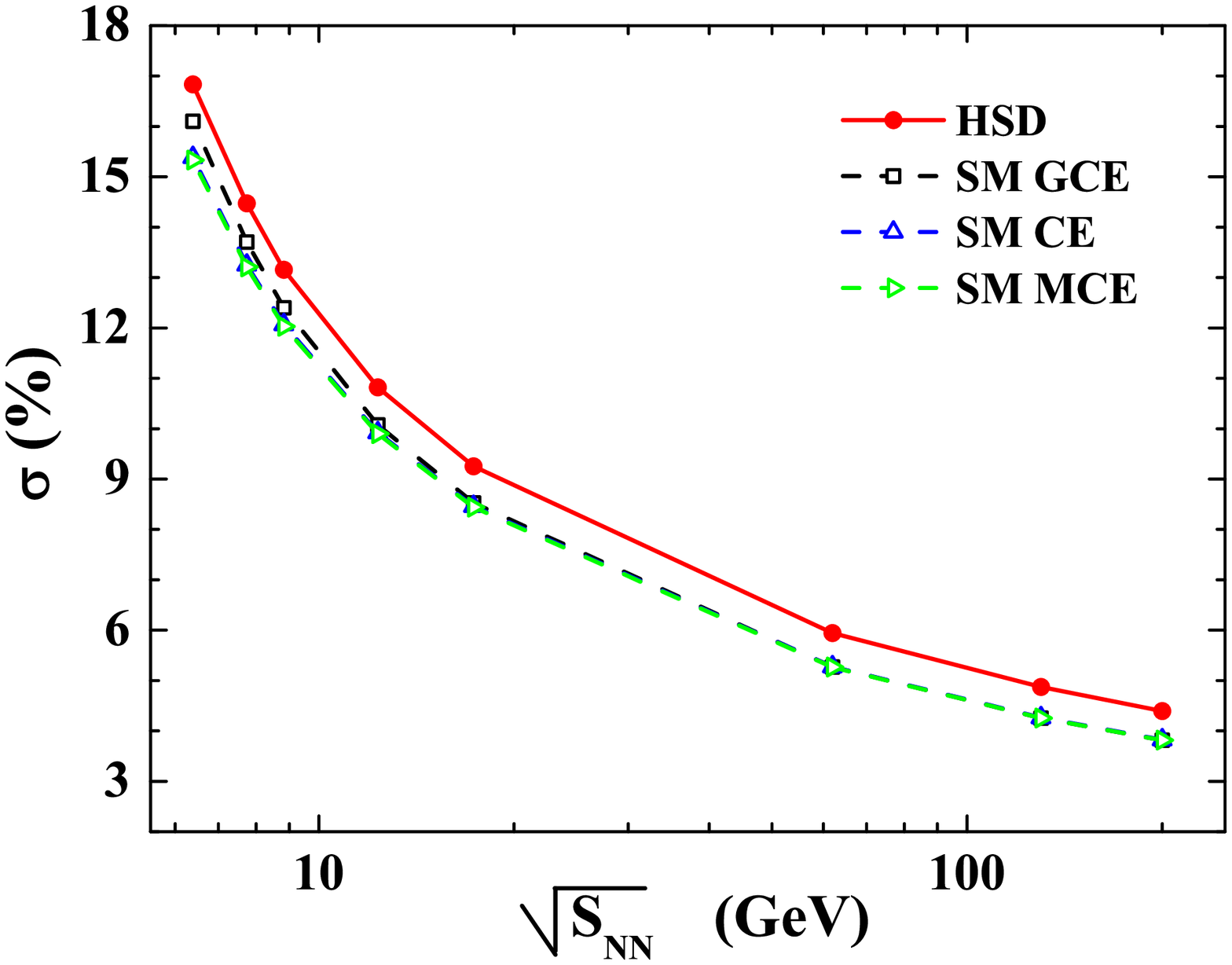,width=8.4cm}
  \epsfig{file=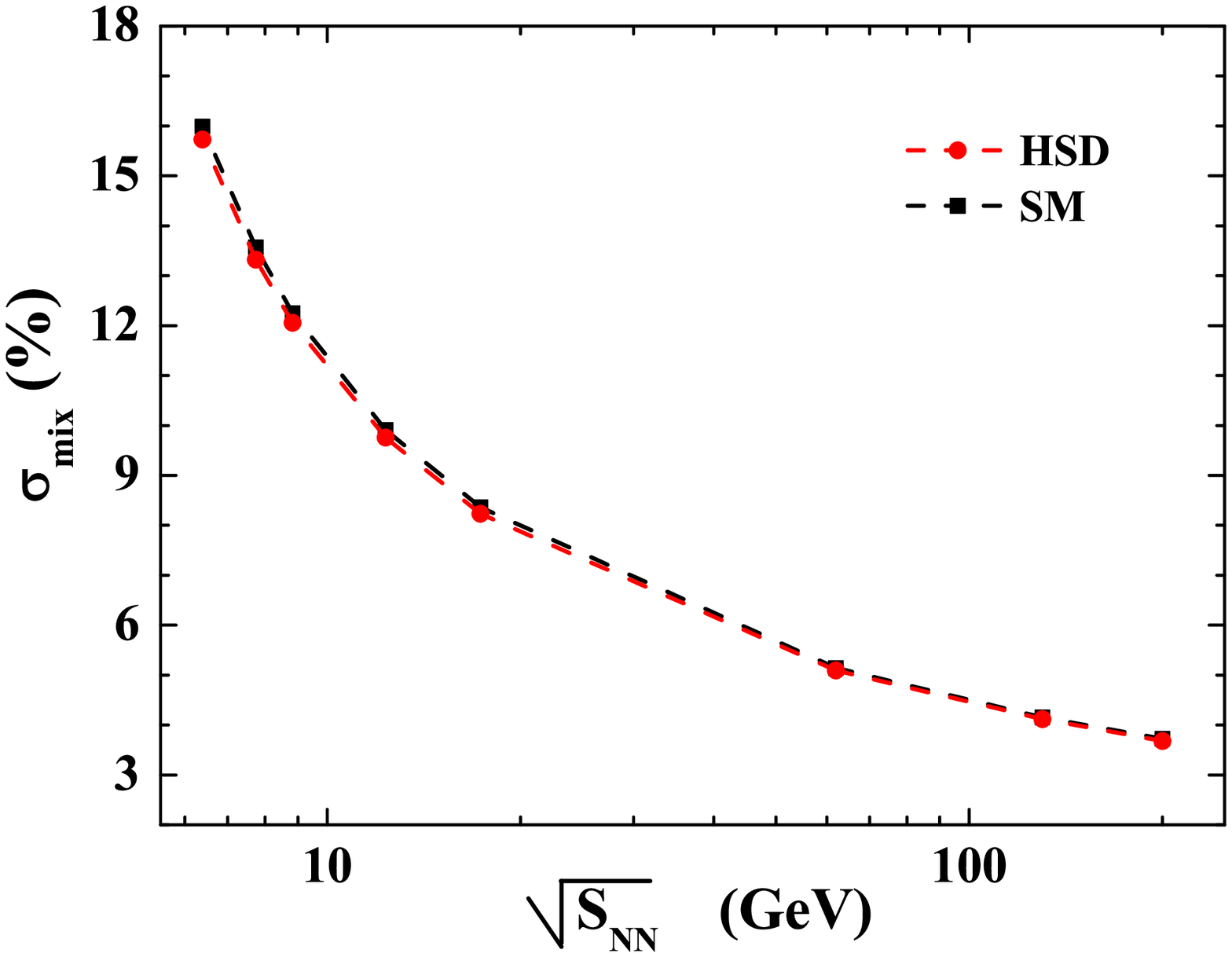,width=8.4cm}
  \caption{(Color online) {\it Left:} The SM results in the GCE, CE, and MCE
  ensembles as well as the HSD results (impact parameter $b=0$) are
presented for $\sigma \cdot$100\% defined by Eq.~(\ref{sigmaKpi}) for
Pb+Pb (Au+Au) collisions  at different c.m. energies $\sqrt{s_{NN}}$.
{\it Right:} The same as in the {\it left} panel, but for $\sigma_{mix}
\cdot$100\% in mixed events defined by Eq.~(\protect\ref{Dmix}),
$\sigma^2_{mix} =1/\langle N_K\rangle +1/\langle N_{\pi}\rangle$.}
\label{fig3}
\end{figure}

In Fig.~\ref{fig3} the values of $\sigma$ (in percent) -- calculated
according to Eq.~(\ref{sigmaKpi}) and Eq.~(\ref{Dmix}) -- are presented
in the {\it left} and {\it right} panel, respectively, for the SM in
different ensembles as well as for the HSD simulations.
The first conclusion from Fig.~\ref{fig3} ({\it left}) is that all
results for $\sigma$ in the different models are rather similar. One
observes a monotonic decrease of $\sigma$ with collision energy. This
is just because of an increase of the kaon and pion average multiplicities
with collision energy. The mixed event fluctuations $\sigma_{mix}$ in
the model analysis are fully defined by these average
multiplicities according to Eq.~(\ref{Dmix}). The values of
$\sigma_{mix}$ are therefore the same in the different statistical
ensembles. They are also very close to the HSD values because we have
fixed the statistical system volume to obtain the same kaon average
multiplicities in the statistical model as in HSD at each collision
energy. As seen from Fig.~\ref{fig3} ({\it right}) the requirement of
$\langle N_K\rangle_{stat}= \langle N_K\rangle_{HSD}$ leads to practically
equal values of $\sigma_{mix}$ in both HSD and the SM.
\begin{figure}[ht!]
  \epsfig{file=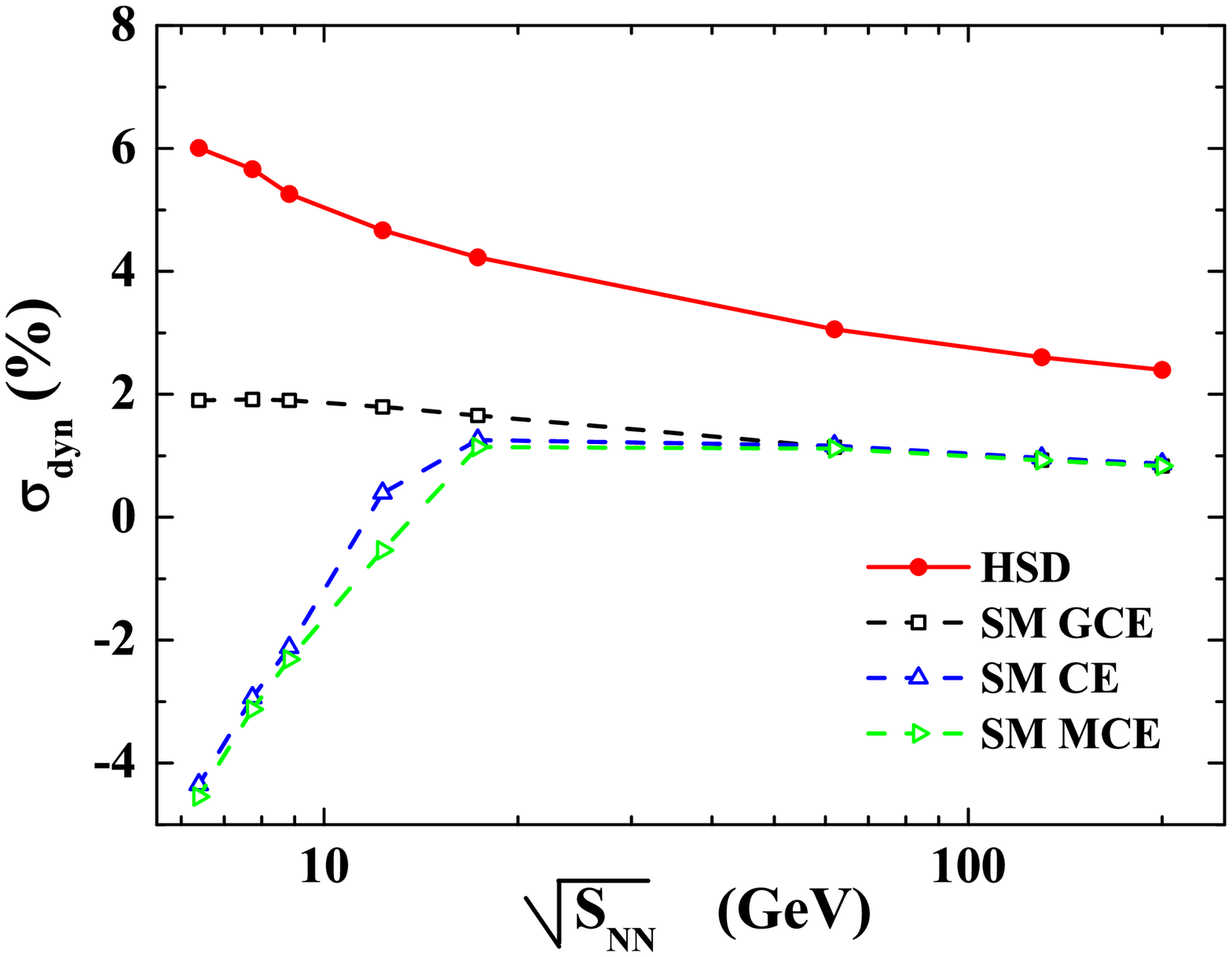,width=8.4cm}
  \epsfig{file=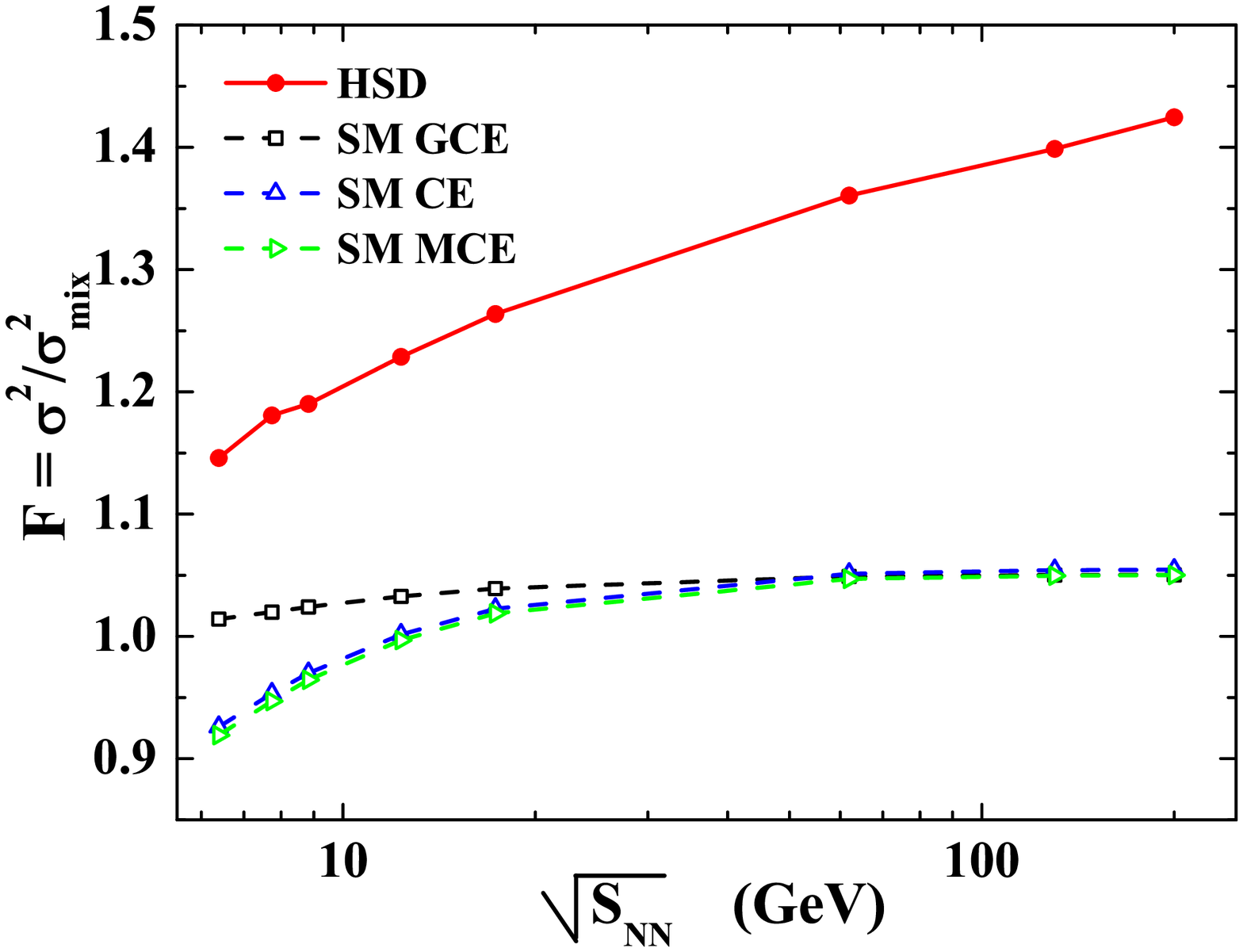,width=8.4cm}
  \caption{(Color online) {\it Left}: The results for the $K/\pi$ fluctuations at
different c.m. energies $\sqrt{s_{NN}}$ in the GCE, CE, and MCE
ensembles as well as from HSD (impact parameter $b=0$) are presented
for $\sigma_{dyn}\cdot$100\% defined by Eq.~(\protect\ref{sigmadyn}).
{\it Right:} The same as in the {\it left} panel but for
$F=\sigma^2/\sigma_{mix}^2$.}
\label{fig4}
\end{figure}

Differences between the statistical ensembles as well as between
the statistical and HSD results become visible for other measures
of $K/\pi$ fluctuations such as $\sigma_{dyn}$ defined by
Eq.~(\ref{sigmadyn}) and $F=\sigma^2/\sigma_{mix}^2$. They are
shown in Fig.~\ref{fig4}, {\it left} and {\it right},
respectively. At small collision energies the CE and MCE results
in Fig.~\ref{fig4} demonstrate negative values of $\sigma_{dyn}$,
respectively $F<1$. When the collision energy increases,
$\sigma_{dyn}$ in the CE and MCE ensembles becomes positive, i.e.
$F>1$. Moreover, the different statistical ensembles approach to
the same values of $\sigma_{dyn}$ and $F$ at high collision energy.
In the SM the values of $\sigma$ and $\sigma_{mix}$
approach zero at high collision energies due to an increase of
the average multiplicities. The same limit should be also valid
for $\sigma_{dyn}$ in the SM. In contrast, the measure $F$ shows a
different behavior at high energies: the SM gives $F\cong 1.05$ in
the high energy limit, while the HSD result for $F$ demonstrates a
monotonic increase with collision energy.
An interesting feature of the SM is approximately the same result for
$\sigma$ (and, thus, $\sigma_{dyn}$ and $F$) in the CE and MCE
ensembles. From Table I and Figs.~\ref{fig1} and \ref{fig2} one
observes that both $\omega_K$, $\omega_{\pi}$ and $\rho_{K\pi}$ are
rather different in the CE and MCE. Thus, as discussed above, an exact
energy conservation influences the particle scaled variances and
correlations. These changes are, however, cancelled out in the
fluctuations of the kaon to pion ratio.

\subsection{Volume Fluctuations}

It has been mentioned in the literature (see, e.g., Ref. \cite{BH}) that
the particle number ratio is independent of volume fluctuations since both
multiplicities are proportional to the volume. In fact, the {\it
average multiplicities} $\langle N_K\rangle$ and $\langle
N_{\pi}\rangle$, but not $N_K$ and $N_{\pi}$, are proportional to the
system volume. Let us consider the problem in the SM assuming the
presence of volume fluctuations at fixed values of $T$ and $\mu_B$.
This assumption corresponds approximately to volume fluctuations in
nucleus-nucleus collisions from different impact parameters in each
collision event. Under these assumptions the SM values in Table I
remain the same for any volume (if only this volume is large enough and
the finite size corrections can be neglected). However, the average
hadron multiplicities are proportional to the volume. Therefore, the SM
result for $\sigma^2$ reads,
$\sigma^2= \sigma_0^2 V_0/V,$
where $V_0$ is the average system volume, and $\sigma^2_0$ is
calculated for the average multiplicities corresponding to this average
volume $V_0$. Expanding
$V_0/V=V_0/(V_0+\delta V)$ in powers of $\delta V/V_0$,
one finds to second order in $\delta V/V_0$,
\eq{\label{V3}
\sigma^2 ~\cong~ \sigma_0^2~\left[1~+~\frac{\langle \left(\delta V\right)^2\rangle}{V_0^2}\right]
~,
}
where
\eq{\label{V4}
\langle \left(\delta V\right)^2\rangle~=~\int dV~(V~-~V_0)^2~W(V)~
}
corresponds to an average over the volume distribution function $W(V)$
which describes the volume fluctuations. As clearly seen from
Eq.~(\ref{V3}) the volume fluctuations influence, of course, the
$K/\pi$ particle number fluctuations and make them larger. Comparing
the $K/\pi$ particle number fluctuations in, e.g., 1\% of most central
nucleus-nucleus collisions with those in, e.g., 10\% one should take
into account two effects. First, in the 10\% sample the average volume
$V_0$ is smaller than that in 1\% sample and, thus, $\sigma_0^2$ in
Eq.~(\ref{V3}) is larger. Second, the volume fluctuations (\ref{V4}) in
the 10\% sample is larger, and this gives an additional contribution to
$\sigma^2$ according to Eq.~(\ref{V3}).

One may also consider volume fluctuations at fixed energy and conserved
charges (see, e.g., Ref.~\cite{power}). In this case the connection
between the average multiplicity and the volume becomes more
complicated. The volume fluctuation within the MCE ensemble can
strongly affect the fluctuations in the particle number ratios. This
possibility will be discussed in more detail in a forthcoming study.

\section{Excitation function for the $K/\pi$ ratio: Comparison with data}

\begin{figure}[ht!]
  \epsfig{file=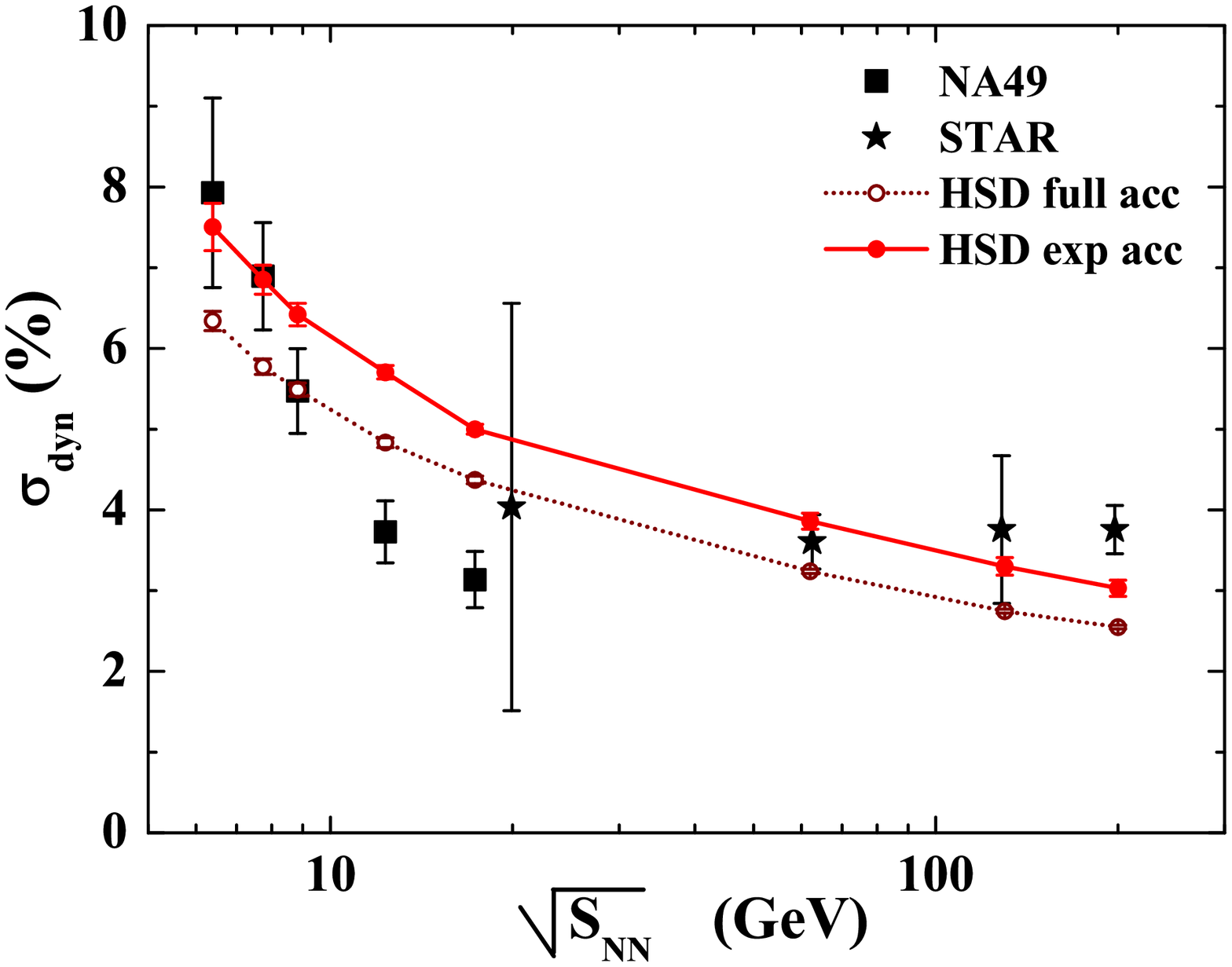,width=8.4cm}
  \epsfig{file=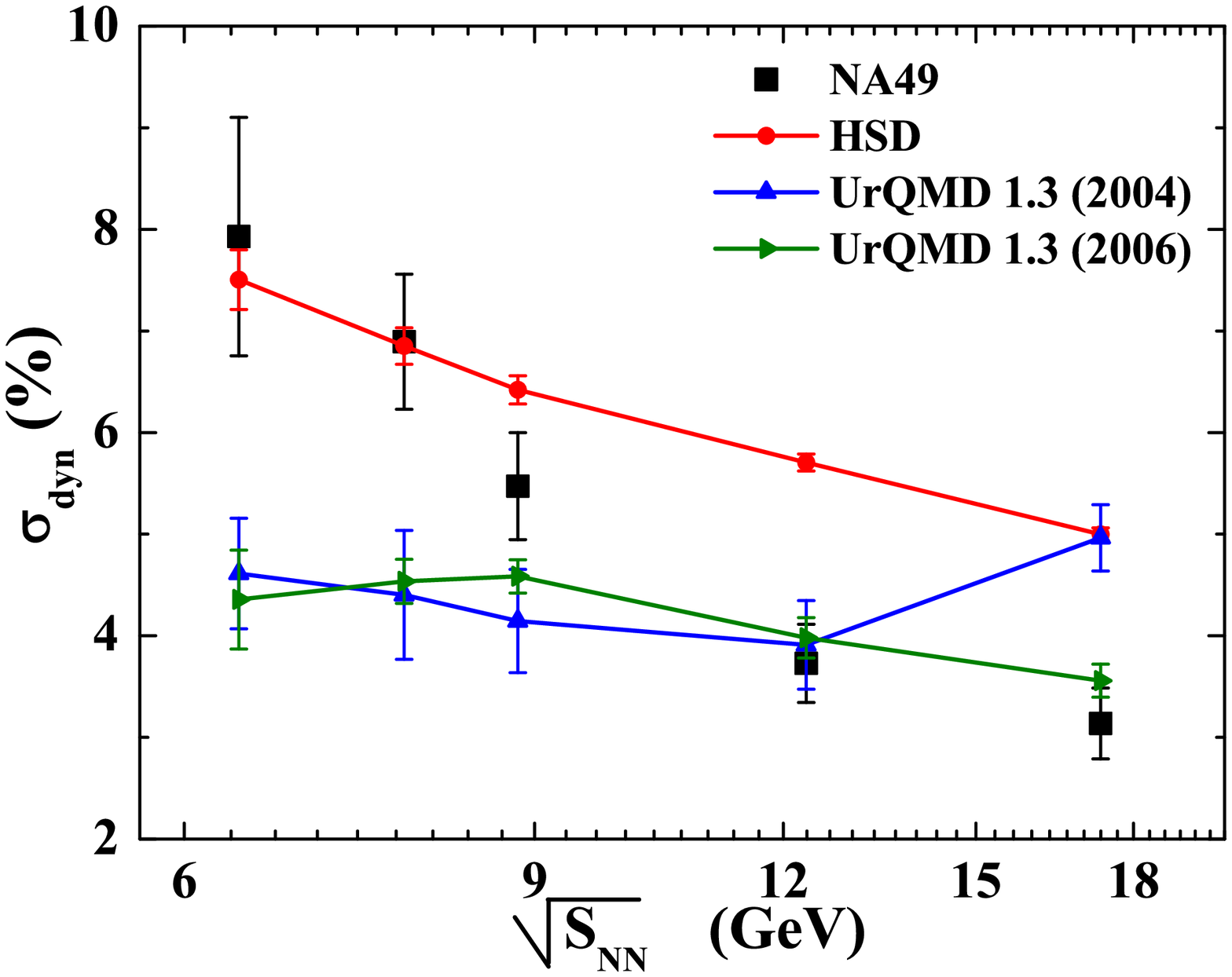,width=8.4cm}
  \caption{(Color online) {\it Left:} The HSD results for the excitation function in
$\sigma_{dyn}$ (\protect\ref{sigmadyn}) for the $K/\pi$ ratio for full
acceptance ( dotted line) and within the experimental acceptance (solid
line) in comparison to the experimental data measured by the NA49
Collaboration at the SPS CERN~\protect\cite{NA49-2} and by the STAR
Collaboration at BNL RHIC~\protect\cite{STAR}. 3.5\% most central HSD
events were selected for the analysis for SPS energies and 5\% -- for
the RHIC energies.  {\it Right:} The HSD results (circles) and two
different versions of UrQMD (triangles) calculations
\protect\cite{Ble,Kresan} for $\sigma_{dyn}$ versus the NA49 data.
Statistical uncertainties in the transport calculations are shown by
error bars. }
\label{fig5}
\end{figure}

A comparison of the SM results for $K/\pi$ fluctuations in
different ensembles with the data looks problematic at present.
This is because of difficulties with implementing  the
experimental acceptance in the SM (see a discussion of this point in
Ref.~\cite{acc}). A similar problems exist in the SM with chemical
non-equilibrium effects discussed in Ref.~\cite{Tor}.
The experimental acceptance can be taken into account in the
transport code. In order to compare the HSD calculations with the
measured data the experimental cuts are applied for the
simulated set of the HSD events.
In Fig. \ref{fig5} the HSD results for the excitation function in
$\sigma_{dyn}$ (\ref{sigmadyn}) for the $K/\pi$ ratio is shown in
comparison with the experimental data measured by the NA49
Collaboration at the SPS CERN~\cite{NA49-2} and by the STAR
Collaboration at BNL RHIC~\cite{STAR}.

For the SPS energies we used a cut $p_{lab}\geq 3$~GeV/c
applied by NA49 to provide a precise particle identification.
For the RHIC energies the cuts are in pseudorapidity, $|\eta|<1$,
and in the transverse momentum, $0.2<p_T<0.6$~GeV/c,~\cite{STAR}.
We note also, that the HSD results presented in Fig.~\ref{fig5}
correspond to the specific centrality selections as in the
experiment - the NA49 data correspond to the 3.5\% most central
collisions selected via the veto calorimeter, whereas in the STAR
experiment the 5\% most central events with the highest
multiplicities in the pseudorapidity range $|\eta|<0.5$ have been
selected.

The HSD results - within the acceptance cuts - are shown in Fig.~\ref{fig5}
({\it left}) as solid lines. In addition in the l.h.s. of Fig.
\ref{fig5} the result for the full acceptance is indicated
by a dotted line.
One can see that the experimental cuts lead to a systematic increase
of $\sigma_{dyn}$, however, do not change the shape of the excitation
function. By comparing the full acceptance line to those in
Fig.~\ref{fig4} ({\it left}) for $b=0$ one sees also a small enhancement of
$\sigma_{dyn}$ which is due to slight decrease of the hadron
multiplicities and, correspondingly, increase of those fluctuations.

In the {\it right} panel of Fig.~\ref{fig5} the HSD results for
$\sigma_{dyn}$ within the experimental acceptance are compared
with two different versions of UrQMD v1.3 simulations
\cite{Ble,Kresan} and the NA49 data in the SPS energy range.
The remaining differences between the UrQMD v1.3 calculations from
2006 and 2004 at 160 A GeV can be attributed to the differences in
implementation of acceptance cuts (cf. discussion in Ref.
\cite{Kresan}). One sees that the UrQMD model gives practically
a constant $\sigma_{dyn}$, which is about $~40 \%$ smaller than
the results from HSD at the lowest SPS energy.  This difference
between the two transport models might be attributed to different
realizations of the string and resonance dynamics in HSD and
UrQMD: in UrQMD the strings decay first to heavy baryonic and
mesonic resonances which only later on decay to `light' hadrons
such as kaons and pions.  In HSD the strings dominantly decay
directly to `light' hadrons (from the pseudoscalar meson octet) or
the vector mesons $\rho$, $\omega$ and $K^*$ (or the baryon octet
and decouplet in case of baryon number $\pm 1$). As discussed
in the previous section, $\sigma_{dyn}$ is indeed very sensitive
to the model details at low bombarding energies: the SM in
different ensembles and the HSD give rather different behavior at
the low SPS energies  (cf. Fig. \ref{fig4}, {\it left}).

While the UrQMD results are available presently only up to
the top SPS energy, the HSD model shows a  good agreement with the
recent STAR data \cite{STAR} (cf. Fig. \ref{fig5}, {\it left}). A
good agreement with the STAR data \cite{STAR08_nu} for $K/\pi$
ratio fluctuations in Cu+Cu at $\sqrt{s_{NN}}=$200~GeV was also
obtained in the Multi-Phase Transport Model (AMPT) \cite{AMPT}.
This is in contrast to the corresponding result from the
Heavy-Ion-Jet-Interaction Generator (HIJING) model \cite{HIJING}
which over-predicts substantially the experimental data
\cite{STAR08_nu}.  The difference has been attributed in Ref.
\cite{STAR08_nu} to an absence of the final re-scattering  in
HIJING which is incorporated in AMPT as well as in HSD.


\section{Summary and conclusions}

We have studied the event-by-event fluctuations of the kaon to
pion number ratio in central Au+Au (or Pb+Pb) collisions from low
SPS up to top RHIC energies within the statistical
hadron-resonance gas model for different statistical ensembles --
the grand canonical ensemble (GCE), canonical ensemble (CE), and
micro-canonical ensemble (MCE) -- and in the
Hadron-String-Dynamics transport approach. We have obtained
substantial differences in the HSD and statistical model results
for the scaled variances $\omega_K$, $\omega_{\pi}$ and the
correlation parameter $\rho_{K\pi}$  as presented in
Figs.~\ref{fig1} and \ref{fig2}.
Thus, the second moments of the multiplicity distributions  may
serve as a good probe for the amount of equilibration achieved in
central nucleus-nucleus collisions. Note that the differences between
the transport and statistical model results for multiplicity
fluctuations and correlations increase with collision energy (see
Refs.~\cite{HSDpp,urqmdpp}). There are also arguments that the behavior
of higher moments of event-by-event multiplicities may serve as an
important signature of the QCD critical point \cite{Step}).

The observable $\sigma_{dyn}$, which characterizes the
fluctuations of the kaon to pion ratio, shows to be rather
sensitive to the details of the model at low collision energies.
The CE and MCE results in Fig.~\ref{fig4} demonstrate negative
values for $\sigma_{dyn}$, while the GCE gives approximately a
constant positive value for $\sigma_{dyn}$. The HSD results
correspond to larger values of $\sigma_{dyn}$ than those in the
GCE statistical model.  They even show an increase at lower SPS
energies.  When the collision energy increases, the quantity
$\sigma_{dyn}$ in the CE and MCE becomes positive. Moreover, the
different statistical ensembles approach to the same values of
$\sigma_{dyn}$ at high collision energy. This is just because the
values of $\sigma$ and $\sigma_{mix}$ approach zero at high
collision energies.  Thus, the same limit equal to zero
should be also valid for $\sigma_{dyn}$ in the statistical models.
On the other hand, the measure $F=\sigma^2/\sigma^2_{mix}$ shows
another behavior at high energies.  The statistical models give a
constant value $F\cong 1.05$ in the high energy limit, while the
HSD results for $F$ demonstrate a monotonic increase with
collision energy.

We find that the HSD model can qualitatively reproduce the
measured excitation function for the $K/\pi$ ratio fluctuations in
central Au+Au (or Pb+Pb) collisions from low SPS up to top RHIC
energies. We have shown that accounting for the experimental
acceptance as well as the centrality selection has a relatively
small influence on $\sigma_{dyn}$ and does not change the shape of
the $\sigma_{dyn}$ excitation function.  We conclude, that the HSD
hadron-string model - which does not have a QGP phase transition
and not explicitly includes the quark and gluon degrees of freedom
- can reproduce qualitatively the experimental excitation
function. In particular, it gives the rise of $\sigma_{dyn}$ with
decreasing bombarding energy. This fact brings us to the
conclusion that the observable enhancement of $\sigma_{dyn}$
at low SPS energies might dominantly signal non-equilibrium
string dynamics rather than a phase transition of hadronic to
partonic matter or the QCD critical point.


\vspace{0.5cm} {\bf Acknowledgements}

We like to thank M.~Bleicher, W.~Cassing, M.~Ga\'zdzicki,
W.~Greiner, C.~H\"ohne, D.~Kresan, M.~Mitrovski, T.~Schuster,
R.~Stock, H.~Str\"obele, G.~Torrieri, S.~Wheaton, and
O.~S.~Zozulya for useful discussions. This work was in part
supported by the Program of Fundamental Researches of the
Department of Physics and Astronomy of National Academy of
Sciences, Ukraine.



\end{document}